\documentclass[12pt]{article}
\usepackage{longtable}
\usepackage[section]{placeins}
\usepackage[top=1in, left=1in, right=1in, bottom=1in]{geometry}
\usepackage[parfill]{parskip}	% Activate to begin paragraphs with an empty line rather than an indent
\usepackage{graphicx}
\usepackage{caption}
\usepackage{enumerate}
\usepackage{subcaption}
\usepackage[export]{adjustbox}
\usepackage{pdfpages}
\usepackage{amssymb}
\usepackage{xcolor}
\usepackage{changepage}
\usepackage{epstopdf}
\usepackage{bbm}
\usepackage{bm}
\usepackage{multicol}
\usepackage{amssymb}
\usepackage{amsmath}
\usepackage{amsfonts}
\DeclareMathAlphabet{\mathbbmsl}{U}{bbm}{m}{sl}

\usepackage{lscape}
\usepackage{rotating}
\usepackage{setspace}
\usepackage{threeparttable}
\usepackage{booktabs}
\usepackage[compact]{titlesec}
\usepackage{lmodern}
\fontfamily{lmtt}\selectfont
\usepackage[T1]{fontenc}

\usepackage[super,comma,sort&compress]{natbib}
\usepackage{hyperref}
\usepackage{titling}
\usepackage{pifont}
\usepackage{soul}
\usepackage[capposition=top]{floatrow}
\newcommand{\subtitle}[1]{%
  \posttitle{%
    \par\end{center}
    \begin{center}\large#1\end{center}
    \vskip0.5em}%
}
\usepackage{amsthm}
\usepackage{amsmath} %maths
\usepackage{multirow}

\usepackage{listings}

\usepackage{mathrsfs}

\newtheorem*{remark*}{Remark}

\usepackage{xpatch}
\makeatletter
\xpatchcmd{\@thm}{\thm@headpunct{.}}{\thm@headpunct{}}{}{}
\makeatother

\usepackage{mathrsfs}

\usepackage{color}
\begin{document}
\pagestyle{plain}

\newtheoremstyle{mystyle}% name
{\topsep}% Space above
{\topsep}% Space below
{\it}% Body font
{}% Indent amount
{\bf}% Theorem head font
{.}%Punctuation after theorem head
{.5em}%Space after theorem head
{}% theorem head spec
\theoremstyle{mystyle}
\newtheorem{assumptionex}{Assumption}
\newenvironment{assumption}
  {\pushQED{\qed}\renewcommand{\qedsymbol}{}\assumptionex}
  {\popQED\endassumptionex}
\newtheorem{assumptionexp}{Assumption}
\newenvironment{assumptionp}
  {\pushQED{\qed}\renewcommand{\qedsymbol}{}\assumptionexp}
  {\popQED\endassumptionexp}
\renewcommand{\theassumptionexp}{\arabic{assumptionexp}$'$}

\newtheorem{assumptionexpp}{Assumption}
\newenvironment{assumptionpp}
  {\pushQED{\qed}\renewcommand{\qedsymbol}{}\assumptionexpp}
  {\popQED\endassumptionexpp}
\renewcommand{\theassumptionexpp}{\arabic{assumptionexpp}$''$}

\newtheorem{assumptionexppp}{Assumption}
\newenvironment{assumptionppp}
  {\pushQED{\qed}\renewcommand{\qedsymbol}{}\assumptionexppp}
  {\popQED\endassumptionexppp}
\renewcommand{\theassumptionexppp}{\arabic{assumptionexppp}$'''$}

\renewcommand{\arraystretch}{1.3}

\newcommand{\argmin}{\mathop{\mathrm{argmin}}}
\makeatletter
\newcommand{\grande}{\bBigg@{2.25}}
\newcommand{\enorme}{\bBigg@{5}}

\newcommand{\blind}{1}

\newcommand{\tit}{\Large Profile Matching for the Generalization and Personalization \\ of Causal Inferences}

\if1\blind

{\title{\tit\thanks{For comments and suggestions, we thank Ambarish Chattopadhyay, Paul Rosenbaum, Donald Rubin, and two anonymous reviewers. 
For graphics, we thank Xavier Alema{\~n}y.
This work was supported through a Patient-Centered
Outcomes Research Institute (PCORI) Project Program Award (ME-2019C1-16172) and grants from the Alfred P. Sloan Foundation (G-2018-10118, G-2020-13946).}}

\author{Eric R. Cohn\thanks{Department of Biostatistics, Harvard School of Public Health, 677 Huntington Avenue, Boston, MA 02115; email: \url{ericcohn@g.harvard.edu}.} \and Jos\'{e} R. Zubizarreta\thanks{Department of Health Care Policy, Biostatistics, and Statistics, Harvard University, 180 A Longwood Avenue, Office 307-D, Boston, MA 02115; email: \url{zubizarreta@hcp.med.harvard.edu}.}
}

\date{} 

\maketitle
}\fi

\if0\blind
\title{ \tit}
\date{} 
\maketitle
\fi

\begin{abstract}
We introduce profile matching, a multivariate matching method for randomized experiments and observational studies that finds the largest possible unweighted samples across multiple treatment groups that are balanced relative to a covariate profile.
This covariate profile can represent a specific population or a target individual, facilitating the generalization and personalization of causal inferences.
For generalization, because the profile often amounts to summary statistics for a target population, profile matching does not \textcolor{black}{always} require accessing individual-level data, which may be unavailable for confidentiality reasons.
For personalization, the profile \textcolor{black}{comprises the characteristics of} a single individual.
Profile matching achieves covariate balance by construction, but unlike existing approaches to matching, it does not require specifying a matching ratio, as this is implicitly optimized for the data.
The method can also be used for the selection of units for study follow-up, and it readily applies to multi-valued treatments with many treatment categories.
We evaluate the performance of profile matching in a simulation study of \textcolor{black}{the} generalization of a randomized trial to a target population.
We further illustrate this method in an exploratory observational study of the relationship between opioid use and mental health outcomes.
We analyze these relationships for three covariate profiles representing: (i) sexual minorities, (ii) the Appalachian United States, and (iii) \textcolor{black}{the characteristics of} a hypothetical vulnerable patient. 
\textcolor{black}{The method can be implemented via the new function \texttt{profmatch} in the \texttt{designmatch} package for R, for which we provide a step-by-step tutorial.}
\end{abstract}

%\vspace*{.3in}
\clearpage
\doublespacing

%\singlespacing
%\pagebreak
%\tableofcontents
%\pagebreak
%\doublespacing

%%%%%%%%%%%%%%%%%%%%
%%%%%%%%%%%%%%%%%%%%
%%%%%%%%%%%%%%%%%%%%

\section*{Introduction}

The randomized experiment is the most reliable method to learn about the effects of treatments \textcolor{black}{because} randomization provides an unequivocal basis for \textcolor{black}{bias control and inference}. \citep{RN1}
However, due to ethical or practical constraints, investigators must often rely on observational studies, where the treatment assignment is unknown. 
A basic principle in the design of an observational study is to approximate as closely as possible the randomized controlled experiment that would have been conducted under ideal circumstances. \citep{dorn1953philosophy, RN2,RN3,RN4,RN5}
One \textcolor{black}{method} that transparently approximates the structure of a randomized experiment in observational studies is matching.
\textcolor{black}{These methods can help align the times of study eligibility assessment and treatment assignment across units, as in risk set matching,\citep{li2001balanced} and can characterize and target specific populations, as we discuss in this paper.}
\textcolor{black}{In addition, in matching the} unit of analysis remains intact, its adjustments are an interpolation of the \textcolor{black}{available} data, and imbalances in the distributions of the observed covariates are \textcolor{black}{made} patent. 
Multivariate matching methods can also facilitate forms of inference akin to randomized experiments and assist \textcolor{black}{in} sensitivity analyses to hidden biases.\citep{RN6} 
See Rubin (2006),\cite{rubin2006matched} Stuart (2010),\cite{RN8} Imbens (2015),\cite{RN9} and Rosenbaum (2020)\cite{RN10} for perspectives on multivariate matching methods in observational studies.

Matching seeks to find similar units across treatment groups such that the covariate distributions in the matched groups are balanced.\citep{RN11}
In empirical studies, units are often matched on the estimated propensity score.\citep{RN12, RN15, RN14}
However, propensity score matching does not guarantee that the resulting matched samples are adequately balanced due to possible misspecification of the propensity score model or small samples.\citep{RN16}
\textcolor{black}{These issues} can be further complicated when the treatment takes on more than two values. 
Additionally, when there is limited overlap in covariate distributions, the clinical or practical meaning of the final matched sample can be unclear if some treated units need to be discarded in order to attain balance.\citep{RN19}

Recent matching techniques that address some of these limitations leverage developments in modern optimization.\citep{RN20, RN21, RN22, parikh2018malts, RN23} 
For example, cardinality matching\citep{RN24} finds the largest matched sample that is balanced relative to covariate balance requirements specified by the investigator. 
However, the matching ratio in cardinality matching is fixed across treatment groups, \textcolor{black}{while} relaxing this restriction can augment the information of the balanced sample.\citep{RN25}. 
Also, the method has not yet been extended to build matched samples for covariate profiles that can represent general populations or specific individuals, which could facilitate the generalization, transportation, and personalization of causal inferences.

Current methods for generalizing and transporting effect estimates to target populations include the g-formula and inverse probability weighting.\citep{RN50}  
These approaches can be combined to form doubly robust estimators,\citep{RN49} and they have recently been extended to analyses of subgroup heterogeneity in target populations.\citep{RobertsonSarahE2021Esei}
See Josey et al. (2020)\cite{RN52} for a weighting calibration approach to transportation with observational data.
While these approaches tend to favor statistical efficiency, matching methods can privilege interpretability and study design.
For instance, appropriately matched samples can retain the advantages of unweighted data, where the absence of unit-specific weights can facilitate the communication of results to broad audiences and outcome analyses \textcolor{black}{that do not readily support weights}.

Along these lines, we propose profile matching, a new multivariate matching method that finds the largest possible self-weighted samples (that is, samples that do not require weights for their analysis)\cite{cochran1977sampling} that are balanced relative to a covariate profile. 
In profile matching, the covariate profile is flexible and can characterize a target population or a single individual.
This facilitates the generalization and personalization of causal inferences in both randomized and observational studies. 
\textcolor{black}{In certain cases,} profile matching does not require accessing individual-level data of the target population (which may be unavailable for confidentiality reasons), \textcolor{black}{and} summary statistics of the target's covariate distribution can be used as the profile.
For personalization, profile matching can balance \textcolor{black}{sample means around the characteristics of a patient of interest.}
Additionally, \textcolor{black}{subject to data constraints,} the method readily extends to multi-valued (\textcolor{black}{$>$} 2) treatments.

\textcolor{black}{In profile matching, the treatment groups are matched for aggregate covariate balance in the spirit of Stuart (2010).\cite{RN8}
After profile matching, various outcome analyses can follow.
For example, one can perform simple graphical displays of the outcomes or conduct further adjustments in augmented estimators using regression methods.}
Like cardinality matching,\citep{RN24} profile matching for balance can also be followed by full matching for homogeneity\citep{RN13} to provide an explicit assignment between units across treatment groups in the spirit of Rosenbaum (1989).\cite{rosenbaum1989optimal} 

In this paper, we evaluate the method in a simulation study of \textcolor{black}{the} generalization of a randomized trial to a target population, and further illustrate it in an exploratory observational study of the relationship between opioid use treatment and mental health outcomes related to suicide and psychological distress.  
In this study, we analyze these relationships for three distinct covariate profiles: (i) sexual minorities, (ii) the Appalachian United States, and (iii) a hypothetical vulnerable patient.  
We provide R code with step-by-step explanations to implement the methods in the eAppendix.

%%%%%%%%%%%%%%%%%%%%
%%%%%%%%%%%%%%%%%%%%
%%%%%%%%%%%%%%%%%%%%

\section*{Framework}

\subsection*{Setup and notation}

For many questions of scientific interest, the population in a randomized or observational study possibly differs from the target population for which one wishes to estimate the average treatment effect. 
Following Dahabreh and Hern\'an (2019)\cite{DahabrehIssaJ2019Eifa} and Dahabreh et al. (2021),\cite{RN59} we consider a population of units who are selected into either a study population or a target population.
We write $S_i = 1$ for the selection of unit $i$ into the study population and $S_i = 0$ for its selection into the target population.
Units from the study population are then randomly sampled into a randomized trial or observational study that collects covariate, treatment, and outcome data.
Specifically, for each unit $i$, we have a vector $\boldsymbol{X}_i$ of $p$ observed covariates and a binary treatment assignment indicator $Z_i$, with $Z_i = 1$ if the unit is assigned to treatment and $Z_i = 0$ if the unit is assigned to control.
Following the potential outcomes framework for causal inference,\citep{neyman1923application, RN29} each unit has \textcolor{black}{two} potential outcomes under treatment and control, $Y_i(1)$ and $Y_i(0)$, respectively, but only one of them is observed: $Y_i := Z_i Y_i(1) + (1 - Z_i) Y_i(0)$.

\subsection*{Generalization, transportation, and personalization}

This sampling setup is general and can characterize the generalization and transportation of causal inferences \textcolor{black}{as described in Dahabreh and Hern\'an (2019)\cite{DahabrehIssaJ2019Eifa} and Dahabreh et al. (2021)\cite{RN59}} and the personalization \textcolor{black}{of such inferences as described here}.
Generalization covers the cases where the entire target population meets the eligibility criteria for inclusion in the study, whereas transportation covers those situations where at least a portion of the target population is not study-eligible. 
As an example of the former, consider the case where a subset of an observational cohort is included in a randomized trial\textcolor{black}{, and} the investigator seeks to generalize effect estimates from the trial to the entire cohort. 
As an example of the latter, consider a researcher who wishes to estimate the average effect of an intervention in one city using data from a randomized or observational study in another city.
Personalization covers the case where all units in the target population have the same covariate values as an individual of interest.
As an example of this, consider an investigator who wishes to estimate the average effect of an intervention for a population of patients with a similar risk profile to that of a specific patient seeking care.

%\subsection*{Estimand, assumptions, identification, and profiles}
\subsection*{Identification}

Using this framework, our goal is to estimate the target average treatment effect\cite{RN30} given by $\textrm{E}\left[Y_i(1) - Y_i(0) | S_i = 0 \right]$, where the expectation is taken over the distribution of the potential outcomes in the target population. 
Our notation thus far implies the Stable Unit Treatment Value Assumption.\citep{RN31} 
Under the assumptions of strongly ignorable \textcolor{black}{study} selection and treatment assignment, the target average treatment effect can be identified from the observed data.\citep{RN50} 
Sufficient assumptions are:
\begin{enumerate}
%\item \emph{Consistency of potential outcomes}. The observed outcome for individuals with treatment $Z_i = z$ is equal to their potential outcome under treatment $Z_i = z$. That is, $Y_i(z) = Y_i$ for every individual with $Z_i = z$.
\item \emph{Conditional mean exchangeability of treatment assignment in the study population}. Conditional on covariates $\boldsymbol{X}_i$, the potential outcome mean is independent of treatment assignment in the study population: $\textrm{E}[Y_i(z) | \boldsymbol{X}_i = \boldsymbol{x}, Z_i = z, S_i = 1] = \textrm{E}[Y_i(z) | \boldsymbol{X}_i = \boldsymbol{x}, S_i = 1]$, $z = 0, 1$. In randomized trials, this is achieved marginally and by design.
\item \emph{Positivity of treatment assignment}. The probability of being assigned to each treatment is positive, given all values of the covariates $\boldsymbol{X}_i$\textcolor{black}{, among those in the study population}. In randomized trials, this is achieved by design.
\item \emph{Conditional mean exchangeability of selection into the study population}. Conditional on covariates $\boldsymbol{X}_i$, the potential outcome mean is independent of selection into the study population: $\textrm{E}[Y_i(z) | \boldsymbol{X}_i = \boldsymbol{x}, S_i = s] = \textrm{E}[Y_i(z) | \boldsymbol{X}_i = \boldsymbol{x}]$, $z = 0, 1$, \textcolor{black}{$s = 0, 1$.}
\item \emph{Positivity of selection into the study population}. The probability of being selected into the study population is positive, given all values of the covariates $\boldsymbol{X}_i$.
\end{enumerate}

Under these assumptions, the mean of the potential outcome $Y_i(z)$ in the target population is identified \textcolor{black}{from} the observed data by $\textrm{E}\left[Y_i(z) | S_i = 0 \right]= \textrm{E}[\textrm{E}[Y_i | \boldsymbol{X}_i = \boldsymbol{x}, S_i = 1, Z_i = z] | S_i = 0]$. 
\textcolor{black}{In the case of personalization, our estimand is equivalent to a conditional average treatment effect (see, e.g., Chapter 12 of Imbens and Rubin 2015). \cite{RN11}}
%The question is for what values of $\boldsymbol{x}$ the above assumptions hold.
In what follows, we characterize the target population by a covariate profile $\boldsymbol{x}^*$;\citep{chattopadhyay2021implied} that is, a vector of $q \geq p$ summary statistics of its observed covariate distribution.
Usually, $\boldsymbol{x}^*$ amounts to the means of the covariates in the target population, but the profile will ideally include higher-order summary statistics to more completely characterize the distribution of covariates in the target population.

%%%%%%%%%%%%%%%%%%%%
%%%%%%%%%%%%%%%%%%%%
%%%%%%%%%%%%%%%%%%%%

\section*{Profile matching}
%\subsection*{Description of the method}

Profile matching is a new method for building optimal self-weighted samples for causal inference.
Self-weighted\textcolor{black}{, or unweighted,} samples are appealing in concept and in practice.
They have a simple and intuitive structure, as each unit is assigned the same weight (often, equal to one) and they can be used to represent target populations of special interest. \citep{cochran1977sampling}
This enables simple outcome analyses such as graphical displays and more sophisticated prediction approaches that do not readily support unit-specific weights.
In some settings, this also facilitates the selection of units from a larger reservoir for randomization or study follow-up.\citep{stuart2010matching}

Profile matching finds, from a reservoir for each treatment group, the largest possible self-weighted sample that is balanced relative to a target covariate profile.
Profile matching accomplishes this by solving an optimization problem (specifically, an integer programming problem) that maximizes the sample size for analysis subject to certain covariate balance or representativeness constraints.
These constraints are flexible and are determined by the investigator.
They can range from marginal mean balance to joint distributional balance subject to the available data (see Part II of Rosenbaum 2010\cite{RN4} and Resa and Zubizarreta 2016\cite{RN35} for details).
In randomized or observational studies, this facilitates estimation of the target average treatment effect by selecting the covariate profile appropriately.
Profile matching can be implemented directly via a new approach to matching (specifically, by solving a multidimensional knapsack problem)\cite{RN34, RN27} or indirectly via existing software for cardinality matching.\citep{RN24, RN26}
In the eAppendix we describe the technical details of these two approaches.

\begin{sidewaysfigure}[H]
\caption{Schematic of \textcolor{black}{profile matching} for balancing samples toward various targets}
\vspace{0pt}
\includegraphics[scale=1.0]{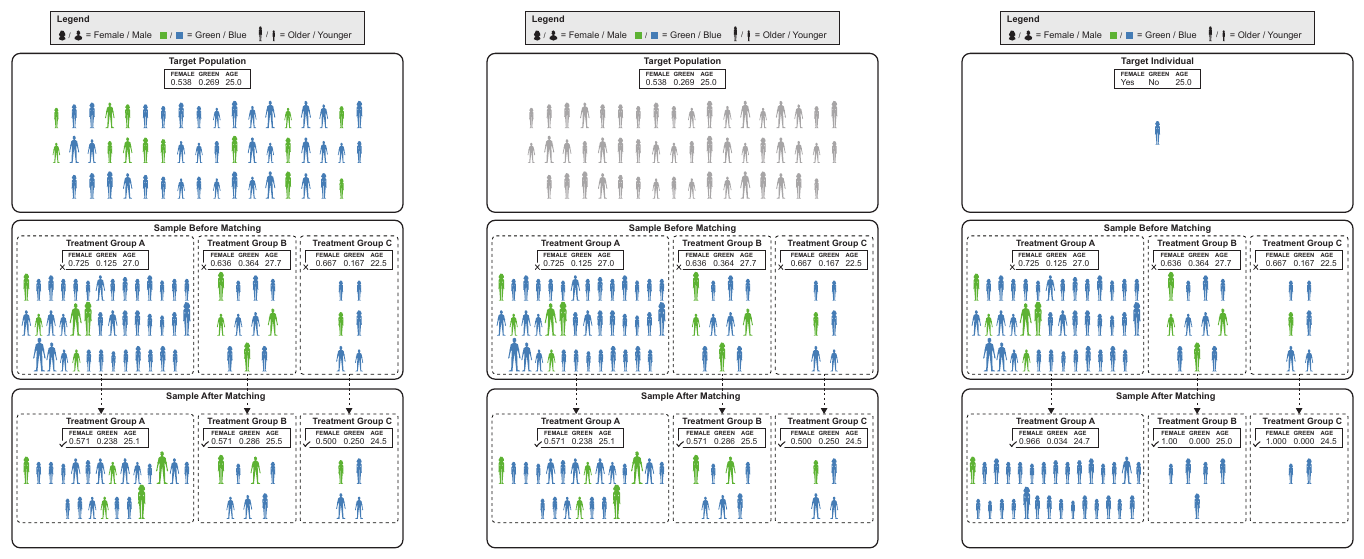}
\vspace{0pt}

\flushleft{\footnotesize{The left panel displays a schematic of profile matching when balancing toward a target population. The target population is summarized by a profile that describes the proportion female, the proportion green, and the average age in the population. The sample before matching is divided into three treatment groups---A, B, and C---described by similar profiles. The groups are both dissimilar from the target and dissimilar from each other. The sample after matching comprises the maximally-sized subsets of each treatment group that are balanced relative to the target population, where the tolerated imbalances after matching are specified by the investigator. The middle panel displays a similar schematic of profile matching when balancing toward a target population and individual-level data are not available. The right panel displays a similar schematic of profile matching when balancing toward a target individual. 
}}
\label{fig:Fig1}
\end{sidewaysfigure}

We illustrate the method in Figure \ref{fig:Fig1}.
In the left panel, we present profile matching whereby the investigator seeks to balance three treatment groups (left, middle panel) toward a target population of interest (left, top panel). 
Prior to matching, there are covariate imbalances across the three treatments as evidenced by their profiles in the summary tables. 
Additionally, they are imbalanced relative to the profile of the target population. 
In this implementation of profile matching, we find the largest subsample of each treatment group such that each achieves balance relative to the target. 
In the left panel of Figure \ref{fig:Fig1}, the maximum allowed imbalance is of size 0.5 for the continuous measure (age) and 5 percent for the dichotomous measures (female and green), as evidenced by comparing the profiles in the bottom left panel (i.e., after matching) to the profile in the target (top left). 
We also see that profile matching requires only access to the profile of the target population (top middle). 
Moreover, the profile need not represent a population; it can also summarize an individual (see right panel of Figure \ref{fig:Fig1}). 
\textcolor{black}{Here, the profile-matched samples have covariate means that are similar to the covariate profile comprising the characteristics of the target individual.}
In this way, profile matching can achieve balance relative to target populations or individuals, even when only summary information is available.

\singlespacing
\doublespacing

%%%%%%%%%%%%%%%%%%%%
%%%%%%%%%%%%%%%%%%%%
%%%%%%%%%%%%%%%%%%%%

\section*{Practical considerations}
\color{black}
Two important considerations for any method of covariate adjustment are the covariates (and functions of covariates) for which to adjust and the level of covariate balance after adjustment.
\color{black}
In profile matching, both of these design choices are explicitly controlled by the investigator through the specification of the covariate profile and the imbalance tolerances.
In what follows, we provide practical guidelines for researchers as they think through these design choices.

\color{black}
First and perhaps foremost, the profile should include covariates that are confounders of either the treatment--outcome relationship or the selection--outcome relationship in order to satisfy the exchangeability assumptions in Section 2.3.
\color{black}
Such covariates can be identified from prior studies and subject matter knowledge, and input from domain experts and key stakeholders can help specify the profile and the balancing constraints.
\color{black}
Important covariates to consider are effect modifiers: covariates that are related to the outcome and across which treatment effects differ.
When the distribution of effect modifiers differs across the study and target populations, treatment effect heterogeneity can follow.\citep{DegtiarIrina2021ARoG}
\color{black}
In deciding what covariates to balance, researchers should then include effect modifiers from the relevant literature, as these are critical to fully characterize effects in target populations.
\color{black}
While fully accounting for confounding and heterogeneity can lead to a more complex profile (and thus more balancing constraints), profile matching is optimal in that it maximizes the effective sample size subject to these balancing constraints and the binary constraints on the resulting weights (i.e., that they take on a value of 0 or 1).
\color{black}
Profile matching may also be a valuable method to estimate and communicate effect modification across populations, as target populations across which effects are estimated can be varied simply (i.e., by varying the profile) and the absence of unit-specific weights can facilitate visual displays of heterogeneity.\cite{zubizarreta2013effect}
\color{black}

\textcolor{black}{More formally, the question of what to balance relates to the potential outcome models $\mathrm{E}[Y_i(z) | \bm{X}_i]$ for $z \in \{0, 1\}$ and what covariates (and their functions) determine them.}
For the simple difference-in-means estimator to have asymptotically negligible bias, the covariate functions included in the profile must approximate well these potential outcome models.
For example, balancing covariate means only is sufficient to remove bias when the outcome model is linear in the covariates.
Balancing higher order sample moments (e.g., all second-order moments, including two-way interactions) is sufficient when the outcome model is linear in second-order transformations of the covariates (e.g., linear in the squares of covariates and in their two-way interactions).
See, e.g., Ben-Michael et al. (2021)\cite{benmichael2021balancing} for additional discussion of the relationship between covariate balance and assumptions about the outcome model.

\textcolor{black}{
The other important consideration for adjustment methods is how well to balance covariates, where in profile matching, this balance is with respect to the profile.}
Ultimately, there is a trade-off between covariate balance and sample size in the profile-matched sample.
Sometimes this decision is made on the basis of substantive knowledge.
For example, it might be medically meaningful to require that the matched samples differ by no more than one year from the value of age in the profile.
Other times, the maximal imbalance in any covariate between two matched samples is set to 0.1 absolute standardized differences in means (ASDMs).
%With profile matching, this involves setting the maximal imbalance with respect to the profile to 0.05 ASDMs so that the resulting difference between any pair of profile matched-samples used to estimate a causal contrast does not exceed the 0.1 threshold.
With simple weighting estimators, however, different thresholds can produce better estimation accuracy than this threshold.\citep{RN35}
\textcolor{black}{Investigators can also use additional methods (e.g., regression models) after profile matching to adjust for residual imbalances.}
To our knowledge, how to optimally select the degree of covariate balance with matching and weighting methods is an open question in the literature.
See Section 7.3 of Ben-Michael et al. (2021)\cite{benmichael2021balancing} for a discussion of this topic.

%%%%%%%%%%%%%%%%%%%%
%%%%%%%%%%%%%%%%%%%%
%%%%%%%%%%%%%%%%%%%%

\section*{Simulation study}

\subsection*{Data generation processes}

We evaluate the performance of profile matching for the problem of generalization in a simulation study based on Hainmueller. \cite{RN58} 
For the sake of \textcolor{black}{illustration}, we model the case where a randomized trial is nested within a larger cohort for which we would like to generalize the average treatment effect from the trial. 
There are six observed covariates, $X_1, ..., X_6$, distributed as follows: 
$(X_1, X_2, X_3)^\top \sim \textrm{MVN}_3 ( (0, 0, 0)^\top, [(2, 1, -1)^\top,$ $ (1, 1, -0.5)^\top,$ $(-1, -0.5, 1)^\top] )$,
$X_4 \sim \text{Unif}(-3, 3)$,
$X_5 \sim \chi_1^2$,
$X_6 \sim \text{Bernoulli}(0.5)$.
The true model for the probability of selection from this cohort into the trial is given by \textcolor{black}{the probit model: 
$\Phi \{ (X_1 + 2X_2 - 2X_3 - X_4 - 0.5X_5 + X_6)/\sigma \}$ where $\Phi$ is the cumulative distribution function of the standard Normal distribution.}
%$S = I(X_1 + 2 X_2 - 2 X_3 - X_4 - 0.5 X_5 + X_6 + \varepsilon  > 0)$ where $I(x > 0)$ is the indicator function equal to 1 if $x > 0$ and equal to 0 otherwise, and $\varepsilon \sim \textrm{N}(0, \sigma^2)$. 
The variance $\sigma^2$ equals either 30 or 100, corresponding to cases of low and high overlap of covariates across the trial and non-trial groups. 
This results in $n_{\textrm{trial}}$ individuals selected into the trial. 
For this group, we assume that the binary treatment $Z_i$ is marginally randomized, such that $\Pr(Z_i = 1) = 0.5$ for all observations $i = 1, ..., n_{\textrm{trial}}$. 
\textcolor{black}{Across all scenarios, $n_{\text{cohort}} = 1500$ and $n_{\textrm{trial}} \approx 750$.} 
For all settings, the true average treatment effect is 0.
 
We implement three different continuous potential outcome models \textcolor{black}{(OM) across two effect heterogeneity settings (A and B)}:

\singlespacing
\begin{enumerate}[OM 1:]
\item 
\begin{enumerate}[A:]
\item $Y(0) = X_1 + X_2 + X_3 - X_4 + X_5 + X_6 + \eta_0$;\\
$Y(1) = X_1 + X_2 + X_3 - X_4 + X_5 + X_6 + \eta_1$
\item $Y(0) = X_1 + X_2 + X_3 - X_4 + X_5 + X_6 + \eta_0$;\\
$Y(1) = X_1 + X_2 + X_3 - X_4 + X_5 + X_6 +  \textcolor{black}{10 \times (X_6 - E[X_6])} + \eta_1$
\end{enumerate}
\item
\vspace{0.1cm}
\begin{enumerate}[A:]
\item $Y(0) = X_1 + X_2 + 0.2 X_3 X_4 - \sqrt{X_5} + \eta_0$\\
$Y(1) = X_1 + X_2 + 0.2 X_3 X_4 - \sqrt{X_5} + \eta_1$
\item $Y(0) = X_1 + X_2 + 0.2 X_3 X_4 - \sqrt{X_5} + \eta_0$\\
$Y(1) = X_1 + X_2 + 0.2 X_3 X_4 - \sqrt{X_5}  + \textcolor{black}{10 \times (X_6 - E[X_6])} + \eta_1$
\end{enumerate}
\item
\vspace{0.1cm}
\begin{enumerate}[A:]
\item $Y(0) = (X_1 + X_2 + X_5)^2 + \eta_0$\\
$Y(1) = (X_1 + X_2 + X_5)^2 + \eta_1$
\item $Y(0) = (X_1 + X_2 + X_5)^2 + \eta_0$\\
$Y(1) = (X_1 + X_2 + X_5)^2  + \textcolor{black}{10 \times (X_6 - E[X_6])} + \eta_1$
\end{enumerate}
\end{enumerate}
\doublespacing
\vspace{0.1cm}

where \textcolor{black}{$\eta_0 \sim \textrm{N}(0, 1)$ and $\eta_1 \sim \textrm{N}(0, 1)$}.
These outcome models are increasing in their levels of nonlinearity and are correlated with the true propensity model. 
\textcolor{black}{Additionally, while models in heterogeneity setting A exhibit no heterogeneity, models in heterogeneity setting B exhibit heterogeneity by $X_6$, where the average treatment effect among those with $X_6 = 0$ is $5$ and among those with $X_6 = 1$ is $-5$, yet the (marginal) average treatment effect is 0.}

\textcolor{black}{
We evaluate profile matching alongside inverse odds weighting using simple and augmented estimators as described in Dahabreh et al (2020).\cite{RN51}
While the inverse odds weighting methods have well-established statistical properties,\citep{DahabrehIssaJ2019Gcif} similar results for profile matching remain a new and open area of research.
In this simulation study, the first profile matching estimator is a simple difference-in-means estimator for the profile-matched samples.
The profile matching augmented estimator is constructed by fitting linear outcome models to each of the profile-matched samples and then predicting the outcomes for the combined treatment or control profile-matched sample.
The estimated treatment effect is then taken as the difference in the mean predicted values.}
The profile matching designs for the probability of selection (PS) vary in their degree of misspecification. They are:

\singlespacing
\begin{enumerate}[PS 1:]
\item Mean balance of $X_1, ..., X_6$
\item Mean balance of $X_1^2, X_2^2, X_3, X_4^2, X_5^2, X_6$
\item Mean balance of $X_1 X_3, X_2^2, X_4, X_5, X_6$
\end{enumerate}
\doublespacing
\vspace{0.1cm}

\textcolor{black}{The inverse odds weighting methods involve fitting two \textcolor{black}{probit} regression models: one for the probability of selection into the study sample, and another for the probability of treatment assignment.
These models are then used to derive the weights and construct normalized inverse odds weighting and doubly robust estimators as in equations 3 and 5 of Dahabreh et al (2020).\cite{RN51}}
\textcolor{black}{With inverse odds weighting methods, we specify probability of selection (PS) models that vary in their degree of misspecification. They are}:

\singlespacing
\begin{enumerate}[PS 1:]
\item $\widehat{\Pr}(S = 1 | X_1, ..., X_6) = \textcolor{black}{\Phi}(\widehat{\beta}_1 X_1 + \widehat{\beta}_2 X_2 + \widehat{\beta}_3 X_3 + \widehat{\beta}_4 X_4 + \widehat{\beta}_5 X_5 + \widehat{\beta}_6 X_6)$
\item $\widehat{\Pr}(S = 1 | X_1, ..., X_6) =\textcolor{black}{\Phi}(\widehat{\beta}_1 X_1^2 + \widehat{\beta}_2 X_2^2 + \widehat{\beta}_3 X_3 +\widehat{\beta}_4 X_4^2 + \widehat{\beta}_5 X_5^2 + \widehat{\beta}_6 X_6)$
\item $\widehat{\Pr}(S = 1 | X_1, ..., X_6) = \textcolor{black}{\Phi}(\widehat{\beta}_1 X_1X_3 + \widehat{\beta}_2 X_2^2 + \widehat{\beta}_3 X_4 + \widehat{\beta}_4 X_5 + \widehat{\beta}_5 X_6)$
\end{enumerate}
\doublespacing
\vspace{0.1cm}

For both the profile matching and the inverse odds weighting designs, PS 1 is correctly specified, PS 2 is slightly misspecified, and PS 3 is heavily misspecified.
\textcolor{black}{Here, both approaches correctly model the probability of treatment.}

\subsection*{Covariate balance and effective sample size}
\singlespacing
\begin{table}
\color{black}
\centering
\caption{Mean target absolute standardized mean differences \textcolor{black}{for design 1} under no effect heterogeneity (A)}
\vspace{-10pt}
\label{tab:Tab1}
\begin{tabular}{lllllllll}
           &         & \multicolumn{3}{c}{High Overlap} &  & \multicolumn{3}{c}{Low Overlap} \\ \cline{3-5} \cline{7-9} 
Covariate & Group   & Before     & PM1      & IOW1     &  & Before    & PM1      & IOW1     \\ \hline
$X_1$     & Treated & 0.635      & 0.047    & 0.056    &  & 1.013     & 0.046    & 0.129    \\
          & Control & 0.634      & 0.047    & 0.056    &  & 1.011     & 0.047    & 0.125    \\
$X_2$     & Treated & 0.594      & 0.048    & 0.055    &  & 0.941     & 0.048    & 0.125    \\
          & Control & 0.599      & 0.048    & 0.055    &  & 0.938     & 0.048    & 0.122    \\
$X_3$     & Treated & 0.597      & 0.048    & 0.056    &  & 0.940     & 0.048    & 0.123    \\
          & Control & 0.592      & 0.048    & 0.056    &  & 0.943     & 0.048    & 0.121    \\
$X_4$     & Treated & 0.252      & 0.048    & 0.047    &  & 0.375     & 0.047    & 0.088    \\
          & Control & 0.248      & 0.048    & 0.045    &  & 0.381     & 0.047    & 0.088    \\
$X_5$     & Treated & 0.096      & 0.043    & 0.054    &  & 0.135     & 0.043    & 0.099    \\
          & Control & 0.096      & 0.043    & 0.053    &  & 0.136     & 0.043    & 0.098    \\
$X_6$     & Treated & 0.084      & 0.039    & 0.045    &  & 0.112     & 0.037    & 0.090    \\
          & Control & 0.072      & 0.038    & 0.046    &  & 0.111     & 0.037    & 0.089    \\ \hline
\end{tabular}
\color{black}
\vspace{4pt}
\flushleft{\footnotesize{IOW = inverse odds weighting; PM = profile matching. Accompanying numbers indicate the specification of each method.}}
\end{table}
\doublespacing

First, we evaluate the performance of the inverse odds weighting and profile matching methods in achieving balance relative to the covariate means in the cohort using  the target absolute standardized mean difference.\cite{RN46}
The target absolute standardized mean difference measures the absolute standardized difference between the mean in the sample after adjustment and the mean in the target population. 
Table \ref{tab:Tab1} presents the average target absolute standardized mean differences across \textcolor{black}{5000} replicates under no effect heterogeneity \textcolor{black}{for design 1, which is correctly-specified.}
Please see the eAppendix for similar results \textcolor{black}{in the other settings.}
Often, absolute standardized mean differences smaller than 0.1 are interpreted as good balance. 
\textcolor{black}{target absolute standardized mean differences before adjustment provide an evaluation of how ``far away'' the target profile is from the study sample, and, as expected, these target absolute standardized mean differences under low overlap are much higher than those under high overlap.}
For the sake of this simulation, covariate imbalances were controlled to be smaller than 0.05 target absolute standardized mean differences for profile matching \textcolor{black}{in order to constrain treated-control absolute standardized mean differences to be less than 0.1.}
In the table, we see that this balance criteria is satisfied by construction (i.e., by design) with profile matching.
In each setting, we see that profile matching tends to result in lower imbalances than inverse odds weighting does, particularly in the low overlap settings.

Next, we compute the effective sample sizes of the inverse odds weighting and profile matching methods. 
For this, we use Kish's\cite{kish1965survey} formula: $(\sum_{i=1}^{n_{\textrm{trial}}} w_i)^2 / \sum_{i=1}^{n_{\textrm{trial}}} w_i^2$, where $w_i$ is unit $i$'s weight after adjustment.
After profile matching, $w_i = 1$ for each selected unit, so the effective sample size is simply the number of units in each matched sample.
Table \ref{tab:Tab2} presents the results under no effect heterogeneity, and results under heterogeneity are in the eAppendix.
As expected, the effective sample sizes tend to be larger under high rather than low overlap for both inverse odds weighting and profile matching.
\textcolor{black}{Overall, while inverse odds weighting results in higher effective sample sizes when the selection model includes fewer covariates, profile matching exceeds inverse odds weighting when there are more covariates, particularly when the study and target populations are more dissimilar.}

\singlespacing
\begin{table}[h!]
\centering
\caption{Mean effective sample size across methods under no effect heterogeneity (A)}
\label{tab:Tab2}
\begin{tabular}{lllllll}
             & PM1   & PM2   & PM3   & IOW1  & IOW2  & IOW3  \\ \hline
High Overlap & 413.4 & 468.1 & 638.5 & 401.1 & 503.9 & 684.1 \\
Low Overlap  & 193.9 & 288.4 & 572.7& 149.1 & 280.0 & 614.8 \\ \hline
\end{tabular}
\vspace{18pt}
\flushleft{\footnotesize{IOW = inverse odds weighting; PM = profile matching. Accompanying numbers indicate the specification of each method.}}
\end{table}
\doublespacing

\subsection*{Accuracy and confidence}
We evaluate the performance of the inverse odds weighting, profile matching, and their augmented estimators in terms of mean absolute bias and root mean square error (RMSE; Table \ref{tab:Tab3}), plus coverage probability and average length for bootstrapped confidence intervals (Table \ref{tab:Tab4}) across the simulations. 
\textcolor{black}{The tables in this section include results under no effect heterogeneity (A) and low overlap, and results for other settings, including under effect heterogeneity (B), are included in the eAppendix.}
\textcolor{black}{For both inverse odds weighting and profile matching, bootstrap confidence intervals were generated by resampling the cohort data 500 times and calculating treatment effect estimates using the resampled data.
Confidence intervals were calculated by adding (or subtracting) 1.96 times the bootstrapped standard error to (or from) the actual estimate.}
\textcolor{black}{Generally, we see that profile matching and inverse odds weighting have different but comparable strengths in terms of bias and RMSE.}
\textcolor{black}{Table \ref{tab:Tab3} shows that, for the non-augmented estimators under high and low overlap, profile matching methods exhibit slightly lower mean absolute bias than inverse odds weighting methods do, while inverse odds weighting methods have lower RMSE. 
Augmenting these estimators with an outcome model nearly always improves their performance and results in more similar mean absolute bias and RMSE for profile matching and inverse odds weighting.
Results in the eAppendix show that the pattern for the non-augmented estimators holds when effect heterogeneity is present, however, inverse odds weighting's improvements over profile matching in terms of efficiency are less pronounced.
%As a result, with the efficiency gains from augmentation, profile matching now slightly outperforms IOW both in terms of MAB and RMSE}.
With augmentation, profile matching performs slightly better than inverse odds weighting both in terms of mean absolute bias and RMSE in the settings with heterogeneity}.

%\textcolor{black}{When the estimand is the average treatment effect in a target population, the calculation of bootstrapped confidence intervals for both methods requires access to individual-level data from the target, as the bootstrap procedure involves resampling both from the study sample and from the target sample.
%When only summary-level data on the target are available,  the target sample is not resampled,  and the profile is kept fixed across bootstrap replications of the PM procedure.
%%In this case, inference is limited to the target sample average treatment effect (TSATE): $TSATE = n_{\text{targ}}^{-1} \sum_{i=1}^{n_\text{targ}} Y_i(1) - Y_i(0)$, where $n_{\text{targ}}$ is the number of units in the target sample from which the summary data was calculated.
%In our simulation study, the availability of individual-level data on the target allowed us to benchmark against IOW and to conduct inference for the TATE.}

In Table \ref{tab:Tab4}, we see that both inverse odds weighting and profile matching methods exhibit good coverage under no effect heterogeneity.
\textcolor{black}{Under high and low overlap, confidence intervals for the non-augmented profile matching estimators are nearly always shorter than the non-augmented inverse odds weighting intervals.
On average, profile matching intervals are nearly half the size of their inverse odds weighting counterparts.
With augmentation, this advantage vanishes, and the methods exhibit similar interval lengths in nearly all cases, except for the highly misspecified models.
The eAppendix shows that a similar pattern holds under effect heterogeneity.
Some bootstrapped inverse odds weighting confidence intervals, however, exhibit slightly less than nominal coverage in this setting, particularly for the augmented estimators under outcome model 1.}
We note that the calculation of profile matching bootstrap confidence intervals is computationally intensive, whereas with inverse odds weighting such intervals are faster to compute \textcolor{black}{and closed-form expressions for confidence intervals exist.\citep{RN51}}

In practice, when individual-level data from a sample from the target population are available, one can incorporate uncertainty about the population by bootstrapping the profile, as we did in our study.
When the profile is obtained from a single individual, the profile can be considered fixed.
Careful consideration is needed to incorporate uncertainty when only summary statistics from a sample from the target population are available.
While our simulation study results show adequate coverage, we note that the formal properties of bootstrapped confidence intervals for our two profile matching estimators remain to be studied. 
Works that study the statistical properties of related balancing estimators include Zhao et al. (2019)\cite{ZhaoQingyuan2019Safi} and Wang and Zubizarreta (2021).\cite{WangYixin2019LSPo}

\singlespacing
\begin{sidewaystable}
\begin{center}
\caption{Mean Absolute Bias (MAB) and Root Mean Square Error (RMSE) across methods and settings under no effect heterogeneity (A) \textcolor{black}{and low overlap}}
\label{tab:Tab3}
\begin{tabular}{llrrrrrrrrrrrr}
\hline
\multicolumn{2}{c}{Outcome Model} & PM1  & PM2  & PM3  & IOW1            & IOW2            & IOW3            & aPM1            & aPM2            & aPM3            & aIOW1           & aIOW2           & aIOW3           \\ \hline
OM 1            & MAB             & 0.12 & 0.24 & 0.12 & 0.49            & 0.35            & 0.21            & 0.11            & 0.24            & 0.11            & 0.15            & 0.32            & 0.11            \\
                & RMSE            & 0.15 & 0.30 & 0.15 & \textless{}0.01 & \textless{}0.01 & \textless{}0.01 & \textless{}0.01 & \textless{}0.01 & \textless{}0.01 & \textless{}0.01 & \textless{}0.01 & \textless{}0.01 \\
OM 2            & MAB             & 0.12 & 0.19 & 0.13 & 0.41            & 0.27            & 0.14            & 0.12            & 0.19            & 0.12            & 0.19            & 0.26            & 0.11            \\
                & RMSE            & 0.16 & 0.24 & 0.16 & \textless{}0.01 & \textless{}0.01 & \textless{}0.01 & \textless{}0.01 & \textless{}0.01 & \textless{}0.01 & \textless{}0.01 & \textless{}0.01 & \textless{}0.01 \\
OM 3            & MAB             & 0.97 & 0.70 & 0.71 & 1.75            & 2.52            & 1.82            & 0.98            & 0.67            & 0.62            & 3.65            & 1.27            & 0.87            \\
                & RMSE            & 1.24 & 0.88 & 0.89 & 0.06            & 0.05            & 0.04            & \textless{}0.01 & \textless{}0.01 & 0.01            & 0.09            & 0.01            & 0.03            \\ \hline
\end{tabular}
\end{center}
\vspace{18pt}
\flushleft{\footnotesize{PM = profile matching; aPM = profile matching augmented with an outcome model; IOW = inverse odds weighting; aIOW = inverse odds weighting augmented with an outcome model. Accompanying numbers indicate the specification of each method.}}
\end{sidewaystable}

\begin{sidewaystable}
\begin{center}
\caption{Coverage probability (cov.) and length (len.) of bootstrapped confidence intervals across methods and settings under no effect heterogeneity (A) \textcolor{black}{and low overlap}}
\label{tab:Tab4}
\begin{tabular}{llrrrrrrrrrrrr}
\hline
\multicolumn{2}{c}{Outcome Model} & PM1  & PM2  & PM3  & IOW1 & IOW2  & IOW3 & aPM1 & aPM2 & aPM3 & aIOW1 & aIOW2 & aIOW3 \\ \hline
OM 1            & Cov.            & 0.97 & 0.97 & 0.96 & 0.93 & 0.95  & 0.95 & 0.96 & 0.96 & 0.96 & 0.95  & 0.96  & 0.95  \\
                & Len.            & 0.63 & 1.27 & 0.61 & 2.09 & 1.75  & 1.02 & 0.60 & 1.27 & 0.58 & 0.69  & 2.02  & 0.53  \\
OM 2            & Cov.            & 0.96 & 0.97 & 0.96 & 0.92 & 0.94  & 0.95 & 0.96 & 0.97 & 0.97 & 0.95  & 0.95  & 0.96  \\
                & Len.            & 0.65 & 1.03 & 0.66 & 1.69 & 1.23  & 0.71 & 0.64 & 1.02 & 0.64 & 0.87  & 1.49  & 0.57  \\
OM 3            & Cov.            & 0.96 & 0.96 & 0.95 & 0.94 & 0.97  & 0.96 & 0.96 & 0.95 & 0.95 & 0.95  & 0.95  & 0.95  \\
                & Len.            & 4.86 & 3.62 & 3.51 & 7.29 & 12.96 & 8.61 & 4.89 & 3.41 & 3.01 & 15.95 & 7.92  & 3.97  \\ \hline
\end{tabular}
\end{center}
\vspace{18pt}
\flushleft{\footnotesize{PM = profile matching; aPM = profile matching augmented with an outcome model; IOW = inverse odds weighting; aIOW = inverse odds weighting augmented with an outcome model Accompanying numbers indicate the specification of each method.}}
\end{sidewaystable}
\doublespacing

%%%%%%%%%%%%%%%%%%%%
%%%%%%%%%%%%%%%%%%%%
%%%%%%%%%%%%%%%%%%%%

\section*{Case study}

\subsection*{Data and measurements}

Data for this section come from four consecutive years (2015--2018) of the National Survey on Drug Use and Health (NSDUH), resulting in a total sample size of 171,766 individuals. 
Ethical review was not required as the data are publicly available.
%The NSDUH is administered annually to provide nationally representative information on tobacco, alcohol, and drug use, as well as on mental health and other health issues. 
We use this cross-sectional data set primarily for illustrative purposes, that is, to demonstrate the matching methods, rather than for the purposes of devising a precise portrait of effect estimates (see Samples et al. 2019\cite{RN39} for an extensive analysis using the first three years of this data set). 
For different targets, corresponding to a particular population or a specific individual, we are interested in understanding the relationship between opioid use and psychological distress and suicidal thoughts or behaviors, after adjusting for differences in subjects' observed characteristics. 
As long as the covariates, the exposure, and the outcomes are measured orderly in time and are subject to the assumptions for identification, our estimates can be granted a causal interpretation. 
%In what follows, we describe these three sets of variables.

The exposure, opioid use in the past 12 months, takes on one of three values for each individual: 0 = no opioid use, 1 = opioid use but no misuse, and 2 = opioid misuse. 
We examine two outcomes related to this exposure: one continuous and one binary. 
The continuous outcome ranges from 0 to 24 and indicates the respondent's level of psychological distress over the past 30 days (a higher score indicates more distress). 
The binary outcome (1 = yes, 0 = no) indicates whether the respondent has either seriously thought about suicide, made plans to kill themselves, or attempted to kill themselves in the past 12 months. 
Covariates used for matching include those associated with opioid use or misuse and suicidal behaviors in prior studies and sociodemographic measures.\citep{RN39} 

We specify three covariate profiles (corresponding to sexual minorities,  the Appalachian United States, and a hypothetical vulnerable patient whose past year psychological distress score indicates severe impairment) to demonstrate the flexibility of profile matching to estimate the target average treatment effect. 
We chose sexual minorities in response to increased calls for more investigation into the relationship between opioid use and suicide for this group,\citep{RN40} and we chose the Appalachian United States as a region that has suffered particularly severely from the recent opioid epidemic.\citep{RN41} 
For the latter case, we use external data reports to define the covariate profile.\citep{RN42, RN43} 
Additional details on the exposure, outcome, covariates, and data can be found in the eAppendix.

\subsection*{Balance toward three target covariate profiles}

For the sake of illustration, the target profiles consist of the means of observed covariates.
For each covariate $p$, we define the balance threshold as
$
\delta_p = 0.05 \times \sqrt{(s_{p, 0}^2 + s_{p, 1}^2 + s_{p, 2}^2)/3}
$
where $s_{p,j}$ is the standard deviation of covariate $p$ in exposure group $j = 0, 1, 2$. 
Profile matching results in three samples of maximal size that are balanced both relative to each other and relative to the target.
Balanced samples for the sexual minority population have the following sample sizes: no opioid use = 661, opioid use without misuse = 3,216, and opioid misuse = 358. Balanced samples for the Appalachian target population have the following sample sizes: no opioid use = 23,186, opioid use without misuse = 9,143, and opioid misuse = 895. Balanced samples for the vulnerable patient \textcolor{black}{profile} have the following sample sizes: no opioid use = 219, opioid use without misuse = 115, and opioid misuse = 57. 
In the eAppendix, we describe and illustrate an alternative approach to profile matching that creates pairwise cardinality matched sets of equal sizes for each pairwise exposure group comparison.

\begin{figure}[h!]
\begin{center}
\caption{Balance before and after profile matching toward the sexual minority population. TASMD = target absolute standardized mean difference.}
\label{fig:Fig2}
\includegraphics[scale=0.8]{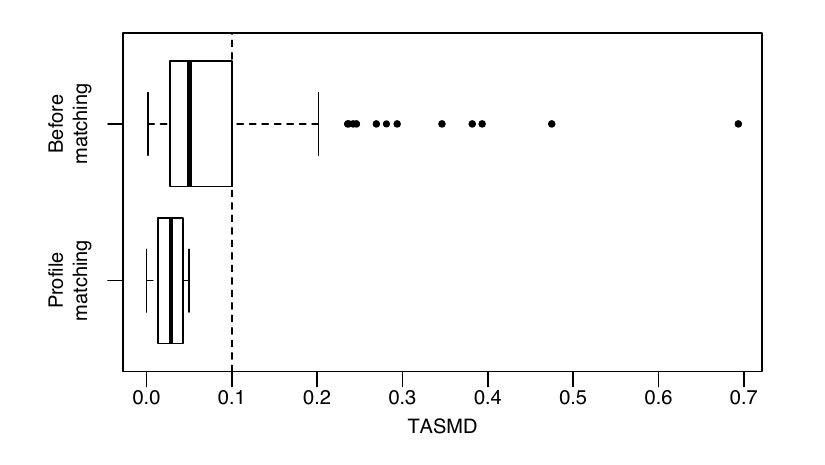}
\end{center}
%\flushleft{\footnotesize{The target average standardized mean differences (TASMDs) are plotted before and after profile matching for balance toward the sexual minority population. The TASMD measures, for each covariate, its difference---appropriately standardized by the pooled standard deviation across the three treatment groups (no opioid use, opioid use without misuse, and opioid misuse in the past year) in the study sample---from the value in the target population: sexual minorities. The samples are from the National Survey on Drug Use and Health 2015-2018.}}
\end{figure}

For the sexual minority covariate profile, we evaluate the performance of profile matching using target absolute standardized mean differences. 
In calculating the target absolute standardized mean difference for each covariate, we use as the denominator the three-way pooled standardized deviation, $\sqrt{(s_{p, 0}^2 + s_{p, 1}^2 + s_{p, 2}^2)/3}$. 
In Figure \ref{fig:Fig2}, we see that profile matching achieves good balance relative to the target population. 
The vertical line signifies a target absolute standardized mean difference equal to 0.1.

\singlespacing
\doublespacing

\begin{figure}[h!]
\begin{center}
\caption{Balance before and after profile matching toward the Appalachian population TASMD = target absolute standardized mean difference.}
\label{fig:Fig3}
\includegraphics[scale=0.75]{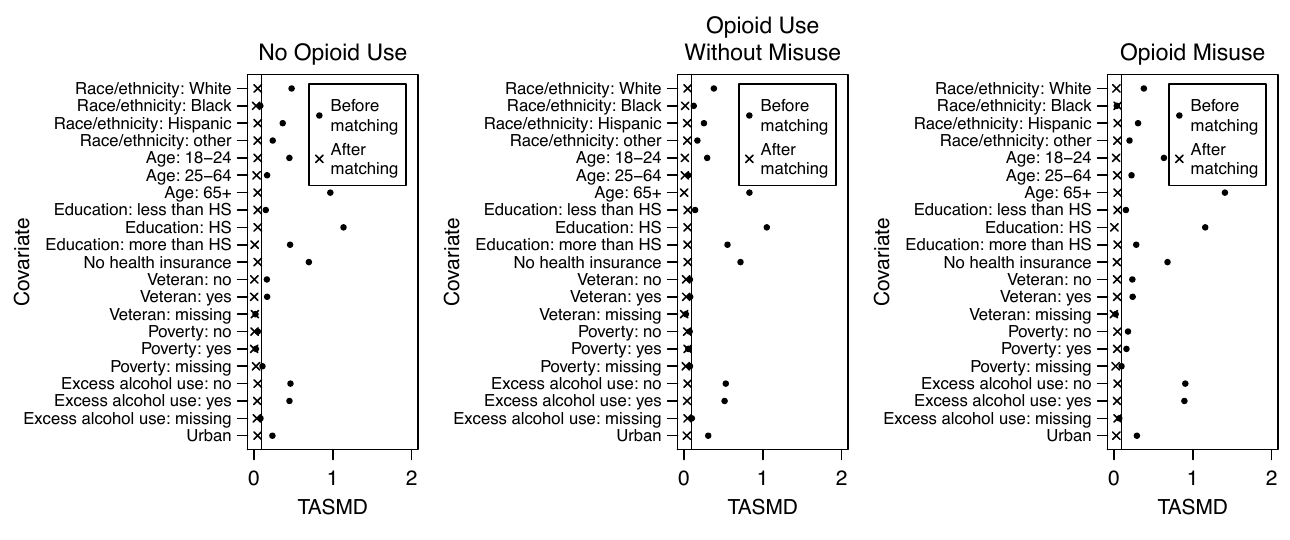}
\end{center}
%\flushleft{\footnotesize{The target average standardized mean differences (TASMDs) are plotted before and after profile matching for balance toward the Appalachian target population. The TASMD measures, for each covariate, its difference—appropriately standardized by the pooled standard deviation across the three treatment groups (no opioid use, opioid use without misuse, and opioid misuse in the past year) in the study sample—from the value in the Appalachian target population. The matched samples are from the National Survey on Drug Use and Health, 2015-2018, and the covariate information for the Appalachian target population comes from the 2012-2016 American Community Survey. HS = high school.}}
\end{figure}

In Figure \ref{fig:Fig3}, we summarize the performance of profile matching for constructing matched sets whose covariate distributions resemble those of the Appalachian target population. Figure \ref{fig:Fig3} plots the target absolute standardized mean differences for each covariate before and after profile matching, for each exposure group. 
The vertical line has a value equal to 0.1. 
Overall, Figure \ref{fig:Fig3} shows that even groups that are relatively imbalanced relative to the target population can be well-balanced using profile matching.

\singlespacing
\doublespacing

For the vulnerable patient \textcolor{black}{profile}, defined as a rural White male between 26 and 34 years old with a high school education whose past year psychological distress score indicates severe impairment, profile matching achieves good balance for each exposure group for all the measured covariates, with target absolute standardized mean differences all well below 0.1 (see eAppendix).

\subsection*{Outcome analyses}

\begin{figure}[h!]
\begin{center}
\caption{Distributions of past month psychological distress by opioid use type after profile matching toward various targets}
\label{fig:Fig5}
\includegraphics[scale=0.82]{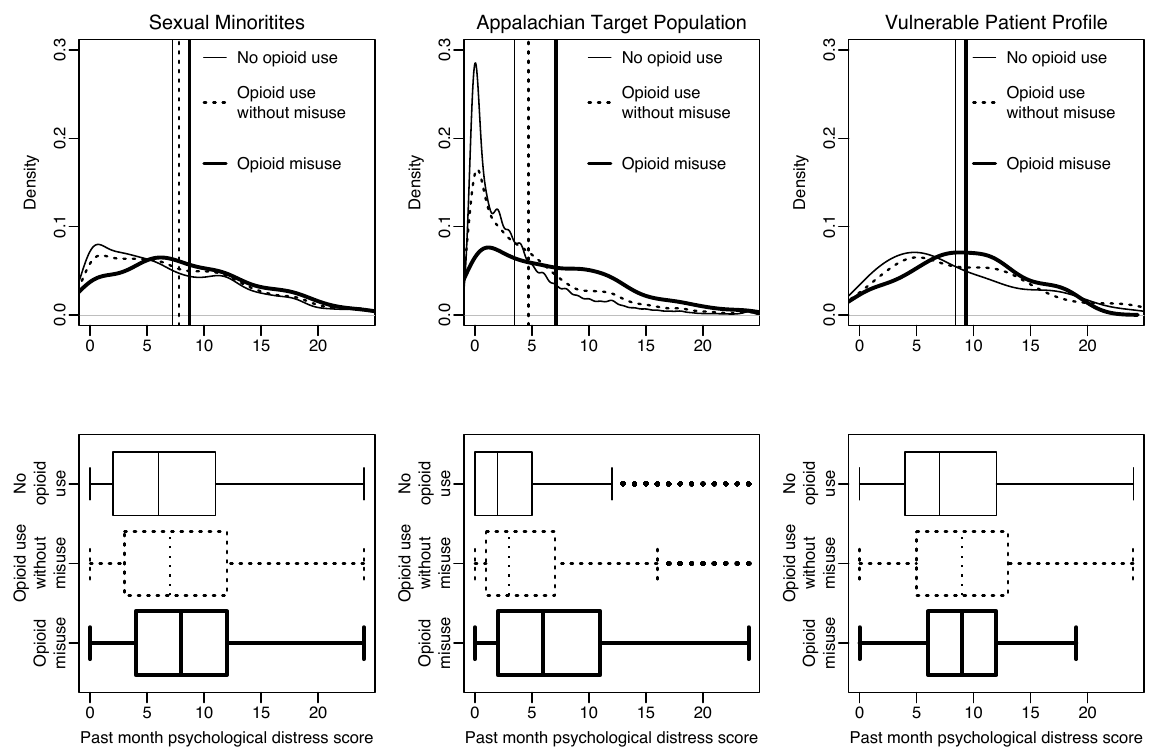} 
\end{center}
%\flushleft{\footnotesize{The distributions of the continuous outcome (past month psychological distress score) are plotted by treatment group (no opioid use, opioid use without misuse, and opioid misuse in the past year) for the three targets: the sexual minority population, the Appalachian target population, and the vulnerable patient. Results reflect profile matching each treatment group for balance toward each target. The top panel plots the probability densities and their means (vertical lines) for each target. The bottom panel displays the results in boxplots.}}
\end{figure}

In Figure \ref{fig:Fig5} we show the distribution of our continuous outcome, the psychological distress score over the past 30 days, for each of the three exposure groups after profile matching for each of the three targets.
In the topmost horizontal panel, the distributions of the outcome for each exposure group and each target population or individual \textcolor{black}{profile} are presented. 
The vertical lines represent the means. 
The lower horizontal panel presents these distributions as boxplots. 
Of note, the means are generally farther to the left than their associated medians, indicating right-skewed distributions whose means are influenced by relatively few observations with high levels of psychological distress. 
Generally, opioid misuse is associated with higher levels of psychological distress, particularly among the Appalachian target population. 
Levels of psychological distress seem similar for those who use but do not misuse opioids compared to those who do not use opioids, and any observed differences are smaller in magnitude than differences in psychological distress between opioid misusers and those who use but do not misuse opioids. 

We present the relationships between the opioid use types and the binary outcome of past year suicidal thoughts, plans, or attempts in Table \ref{tab:Tab5}. For sexual minorities and the Appalachian target population, opioid misuse is associated with a higher probability of past year suicidal thoughts, plans, or attempts relative both to no opioid use and to opioid use without misuse. For these targets, the probabilities of past year suicidal thoughts, plans, or attempts are similar for those who did not use opioids in the past year and those who used but did not misuse. For the vulnerable patient \textcolor{black}{profile}, probabilities are fairly similar across levels of opioid use, perhaps reflecting the fact that this \textcolor{black}{target profile} has a relatively high value for the psychological distress score covariate.

\singlespacing
\begin{table}[h!]
\begin{center}
\caption{Estimated probabilities of past year suicidal thoughts, plans, or attempts by opioid use type after profile matching toward various targets}
\label{tab:Tab5}
\begin{tabular}{lrrr}
\hline
 & \multicolumn{3}{c}{Probability}\\  
Profile matched sample& No opioid use &Opioid use without misuse & Opioid misuse \\ \hline
Sexual minorities & 0.17 & 0.18 & 0.25 \\
Appalachian target population & 0.03 & 0.05 & 0.13 \\
Vulnerable patient & 0.20 & 0.17 & 0.21 \\
\hline
\end{tabular}
\end{center}
\vspace{18pt}
%\flushleft{\footnotesize{The probabilities of past year suicidal thoughts, behaviors or attempts are displayed by treatment group (no opioid use, opioid use without misuse, and opioid misuse in the past year) for the three targets: the sexual minority, the Appalachian target population, and the vulnerable patient. Results reflect profile matching each treatment group for balance toward each target.}}
\end{table}
\doublespacing

%%%%%%%%%%%%%%%%%%%%
%%%%%%%%%%%%%%%%%%%%
%%%%%%%%%%%%%%%%%%%%

\section*{Summary and concluding remarks}

We have proposed a simple yet general method for adjustment in randomized experiments and observational studies.
Profile matching optimally matches units to a description of a population or \textcolor{black}{the characteristics of} a person: the profile.
The method naturally handles multiple treatment groups \textcolor{black}{by solving separate balancing optimization problems for each group relative to a common covariate profile.
By construction, profile matching} maximizes the effective sample size of the matched sample while preserving the unit of analysis and directly balancing the covariates toward the target covariate profile.

%\textcolor{black}{In profile matching, the treatment groups are matched for aggregate covariate balance in the spirit of \cite{RN8}.
%After profile matching, various outcome analyses can follow.
%For example, one can perform simple graphical displays of the outcomes or conduct further adjustments in augmented estimators using regression methods.}
%Akin to cardinality matching \citep{RN24}, profile matching for balance can also be followed by full matching for homogeneity \citep{RN13} to provide an explicit assignment between units across treatment groups in the spirit of \citep{rosenbaum1989optimal}. 

Besides causal inference, profile matching can be used for the selection of units for study follow-up.\citep{stuart2010matching}
Additionally, profile matching can be deployed for hospital quality measurement, where treatments are hospitals\citep{RN47} and template matching samples are replaced by aggregate covariate profiles. 
\textcolor{black}{Finally, this} technique can also be used for the construction of matched samples with evidence factors, where each factor requires a separate balanced matched contrast for a given covariate profile.\citep{RN48}

%\textcolor{black}{Any statistical method such as profile matching has advantages and limitations.
%While profile matching preserves the unit of analysis and prioritizes study interpretability, weighting methods prioritize statistical efficiency by allowing flexible weights.
%Weighting methods also favor computational tractability.
%An interesting direction for future research is to combine profile matching with machine learning adjustments in the form of augmented estimators.}

Any statistical method for adjustment has advantages and disadvantages. While profile matching, in preserving the unit of analysis, can facilitate a simple interpretation of the adjusted sample, weighting methods can often achieve greater statistical efficiency and be more computationally tractable. Also, in the absence of weights, profile matching can be easily followed by more complex adjustment methods from machine learning to form augmented estimators. The properties and performance of such estimators is a promising area of future research.

%%%%%%%%%%%%%%%%%%%%
%%%%%%%%%%%%%%%%%%%%
%%%%%%%%%%%%%%%%%%%%
%\pagebreak
%\onehalfspacing
%\bibliographystyle{asa}
\bibliographystyle{unsrtnat}
\bibliography{mybibliography20}

\newpage
\includepdf[pages=-]{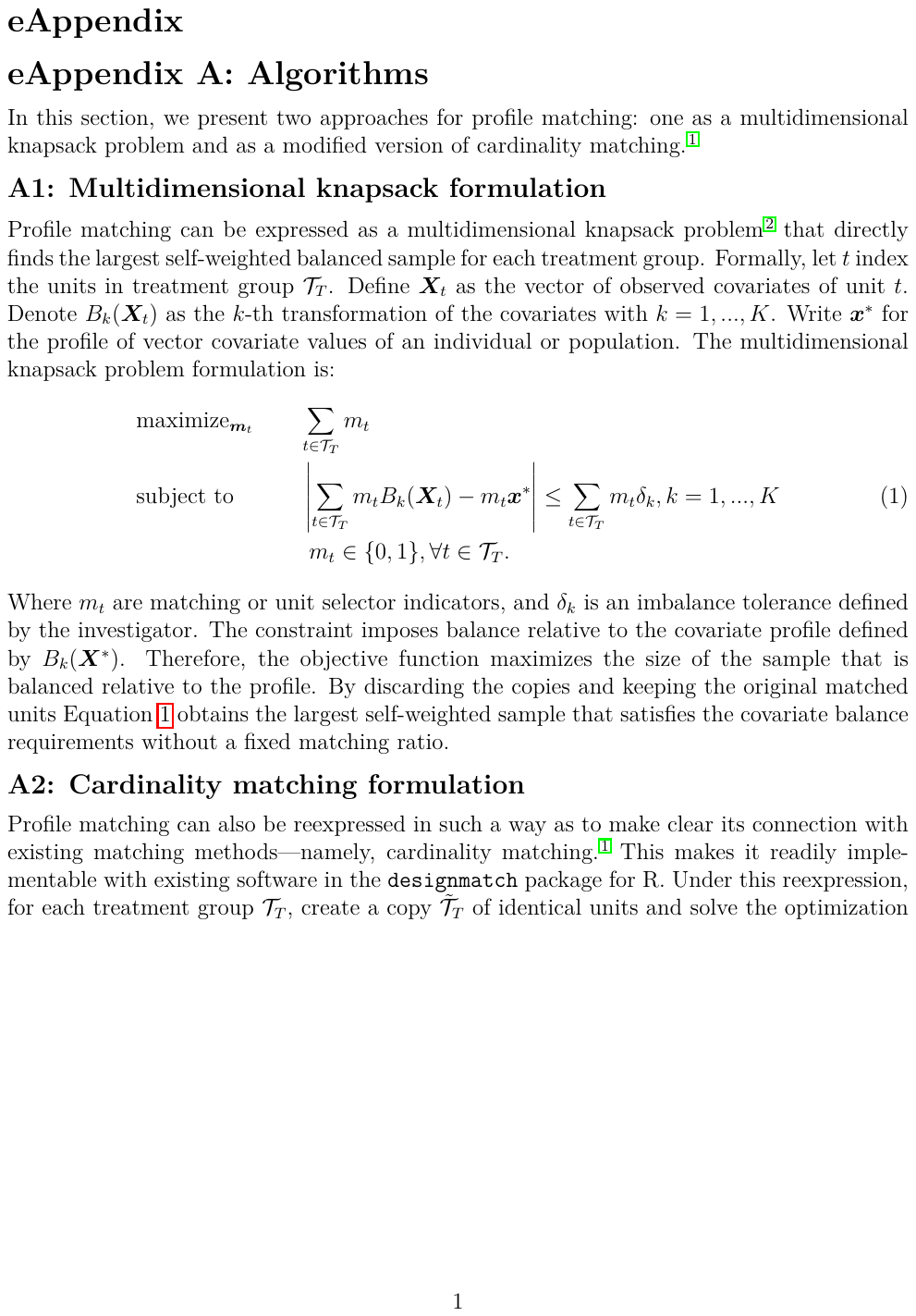}

\end{document}